\documentclass[a4paper,fleqn,usenatbib]{mnras}
\usepackage{newtxtext,newtxmath}

\usepackage{tablefootnote}
\usepackage[T1]{fontenc}
\usepackage{ae,aecompl}
\usepackage[export]{adjustbox}
\usepackage{lipsum}
\usepackage[dvipsnames]{xcolor}

\usepackage{graphicx}	
\usepackage{amsmath}	
\usepackage{amssymb}	

\title[\textit{AstroSat} observation of \textit{MAXI J1535--571}]{Broadband reflection spectroscopy of MAXI J1535--571 using \textit{AstroSat}: Estimation of black hole mass and spin}

\author[N. Sridhar et al.]{
Navin Sridhar$^{1,2}$\thanks{E-mail: navin.sridhar@columbia.edu},
Sudip Bhattacharyya$^{3}$,
Sunil Chandra$^{4}$,
H.M. Antia$^{3}$
\\
$^{1}$Department of Astronomy, Columbia University, New York, NY 10027, USA\\
$^{2}$Department of Physics, Indian Institute of Science Education and Research, Bhauri, Bhopal 462066, India\\
$^{3}$Department of Astronomy and Astrophysics, Tata Institute of Fundamental Research, Mumbai 400005, India\\
$^{4}$Centre for Space Research, North West University, Potchefstroom Campus, Potchefstroom-2520, South Africa\\ 
}

\date{Accepted XXX. Received YYY; in original form ZZZ}

\pubyear{2015}

\begin{document}
\label{firstpage}
\pagerange{\pageref{firstpage}--\pageref{lastpage}}
\maketitle

\begin{abstract}

We report the results from \textit{AstroSat} observations of the transient Galactic black hole X-ray binary MAXI J1535--571 during its hard-intermediate state of the 2017 outburst. We systematically study the individual and joint spectra from two simultaneously observing \textit{AstroSat} X-ray instruments, and probe and measure a number of parameter values of accretion disc, corona and reflection from the disc in the system using models with generally increasing complexities. Using our broadband ($1.3-70$~keV) X-ray spectrum, we clearly show that a soft X-ray instrument, which works below $\sim 10-12$~keV, alone cannot correctly characterize the Comptonizing component from the corona, thus highlighting the importance of broadband spectral analysis. By fitting the reflection spectrum with the latest version of the \textsc{relxill} family of relativistic reflection models, we constrain the black hole's dimensionless spin parameter to be $0.67^{+0.16}_{-0.04}$. We also jointly use the reflection spectral component (\textsc{relxill}) and a general relativistic thin disc component (\texttt{Kerrbb}), and estimate the black hole's mass and distance to be $10.39_{-0.62}^{+0.61}~M_{\odot}$ and $5.4_{-1.1}^{+1.8}$ kpc respectively. 

\end{abstract}




\section{Introduction}\label{Introduction}

The outburst of MAXI J1535--571, a new Galactic X-ray transient, which was soon established to be a black hole X-ray binary (BHXB), was detected with \textit{MAXI}/GSC and \textit{Swift}/BAT in early September 2017 in its hard state \citep{2017ATel10699....1N, 2017ATel10700....1K}. The source was discovered at an X-ray flux level of $34\pm6$ mCrab exhibiting rapid variability, and its brightness was found to be linearly increasing over the course of next few days, until it somewhat levelled off on 2017 September 10, when the source started its transition from the hard state to the soft state \citep{2017ATel10729....1N}. The X-ray detection was quickly followed up with the discovery of an optical counterpart \citep{2017ATel10702....1S}, thus establishing that the source was not only variable in X-rays, but also in the optical/infrared bands. 
Soon after its discovery and when the source was still in hard state, the results of the \textit{NuSTAR} \citep{2013ApJ...770..103H} X-ray data analysis of MAXI J1535--571 were reported by \cite{2018ApJ...852L..34X}. 

They performed an analysis of the reflection spectrum with the \textsc{relxill} family of relativistic reflection models and reported constraints on the parameters such as inner disc radius ($R_\mathrm{in}<2.01~R_\mathrm{ISCO}$; ISCO is the innermost stable circular orbit), black hole spin or Kerr parameter ($a_{\star}>0.84$), coronal lamp-post height ($h=7.2_{-2.0}^{+0.8}~R_\mathrm{g}$; $R_\mathrm{g}$ is the gravitational radius, which is equal to $GM/c^2$, where $M$ is the mass of the black hole), electron temperature ($kT_\mathrm{e}=19.7\pm0.4$ keV) and the absorption column density ($N_\mathrm{H}=8.2_{-0.6}^{+0.3}\times 10^{22}$ cm$^{-2}$).

Follow-up radio observations of the source by \cite{2017ATel10711....1R} on 2017 September 05 revealed a significant radio source at a position consistent with that of optical and X-ray localizations, at flux densities of 7.39 $\pm$ 0.03 mJy and 7.74 $\pm$ 0.05 mJy at 5.5 and 9.0 GHz respectively. With a distance estimate of 6.5 kpc, the high radio luminosity placed the source firmly above what is expected for neutron star X-ray binaries. Within a week, the radio flux saw a dramatic increase, and for the first time, significant sub-mm counterparts were also detected by \cite{2017ATel10745....1T} (220.4$\pm$1.8 mJy at 97 GHz, 226.8$\pm$1.3 mJy at 140 GHz, and 57.7$\pm$1.3 mJy at 230 GHz). Such detections of extremely bright radio and sub-mm counterparts are posited to be arising from a compact synchrotron jet. All these observed properties strongly suggested that the source is a low-mass X-ray binary consisting of a central black hole. 

The onset of the hard to soft state transition of the source by 2017 September 10--11 was not only seen in the form of a steeper spectral index, but also accompanied with the detection of strong quasi-periodic oscillation (QPO) peaks at 1.87 Hz and its harmonic at 3.87 Hz \citep{2018AstL...44..378M}. This transition state was also observed with \textit{NICER}  \citep{2018ApJ...860L..28M} and  \textit{Insight}-HXMT \citep{2018ApJ...866..122H}. Analysis of the \textit{NICER} 
 reflection spectrum of the source by \cite{2018ApJ...860L..28M} led to the constraining of its key parameters, spin of the black hole to its near maximal value ($a_{\star}=0.994(2)$), and the inner truncation of the accretion disc to $1.08(8)~R_\mathrm{ISCO}$. These values are compatible with the earlier results of \cite{2018ApJ...852L..34X} and \cite{2017ATel10768....1G}. While \cite{2018ApJ...852L..34X} had reported distinct disc inclination values ($57^{+1^{\circ}}_{-2}$ and $75^{+2^{\circ}}_{-4}$) for different coronal geometry models, \cite{2018ApJ...860L..28M} reported distinct inclination values ($37^{+22^{\circ}}_{-18}$ and $67.4(8)^{\circ}$) based on the detection of a narrower Fe K line from a different region of the disk. These observations, with different inclination values are indicative of a warped disc structure.
 
 Over the next 10 days, the 2.0--10.0 keV source flux increased from $3\times 10^{-8}$ erg cm$^{-2}$ s$^{-1}$ to $12\times 10^{-8}$ erg cm$^{-2}$ s$^{-1}$, indicating that the spectrum was getting softer \citep{2017ATel10761....1S}. After spending $\sim$2 months in the intermediate states and going through a brief hardening phase, the source finally exhibited its softest spectrum on 2017 November 27 \citep{2017ATel11020....1S}, thus culminating its transition across the intermediate states to the high/soft state. However, that was not immediately followed by any state transition, rather only by an exponential decrease in the brightness of the source, that was observed till 2018 April 30 \citep{2018ATel11568....1N}. After intermittent transitions back and forth between hard and soft states and an unexpected re-brightening event, the source ultimately began its transition to the hard state on 2018 May 26 \citep{2018ATel11682....1N}.

\textit{AstroSat} observation of MAXI J1535--571 was triggered on 2017 September 12, during the onset of the first hard to soft state transition. Here, using two \textit{AstroSat} instruments, we, for the first time, characterize the simultaneous broadband spectrum of the source in the energy range of $1.3-70$~keV. We also perform reflection spectroscopic analysis using the \textit{AstroSat} data, and estimate the black hole's mass, its dimensionless spin parameter, source distance and its luminosity. In section~\ref{obs}, we describe the observation and data reduction with the \textit{AstroSat} X-ray instruments, in section~\ref{results}, we detail the results from our analyses, and in section~\ref{discussion}, we discuss the implications of our results in comparison to the previous works on this source.

\section{Observations and Data  Reduction}\label{obs}

\begin{figure}
	\includegraphics[width=\columnwidth, scale=1.5]{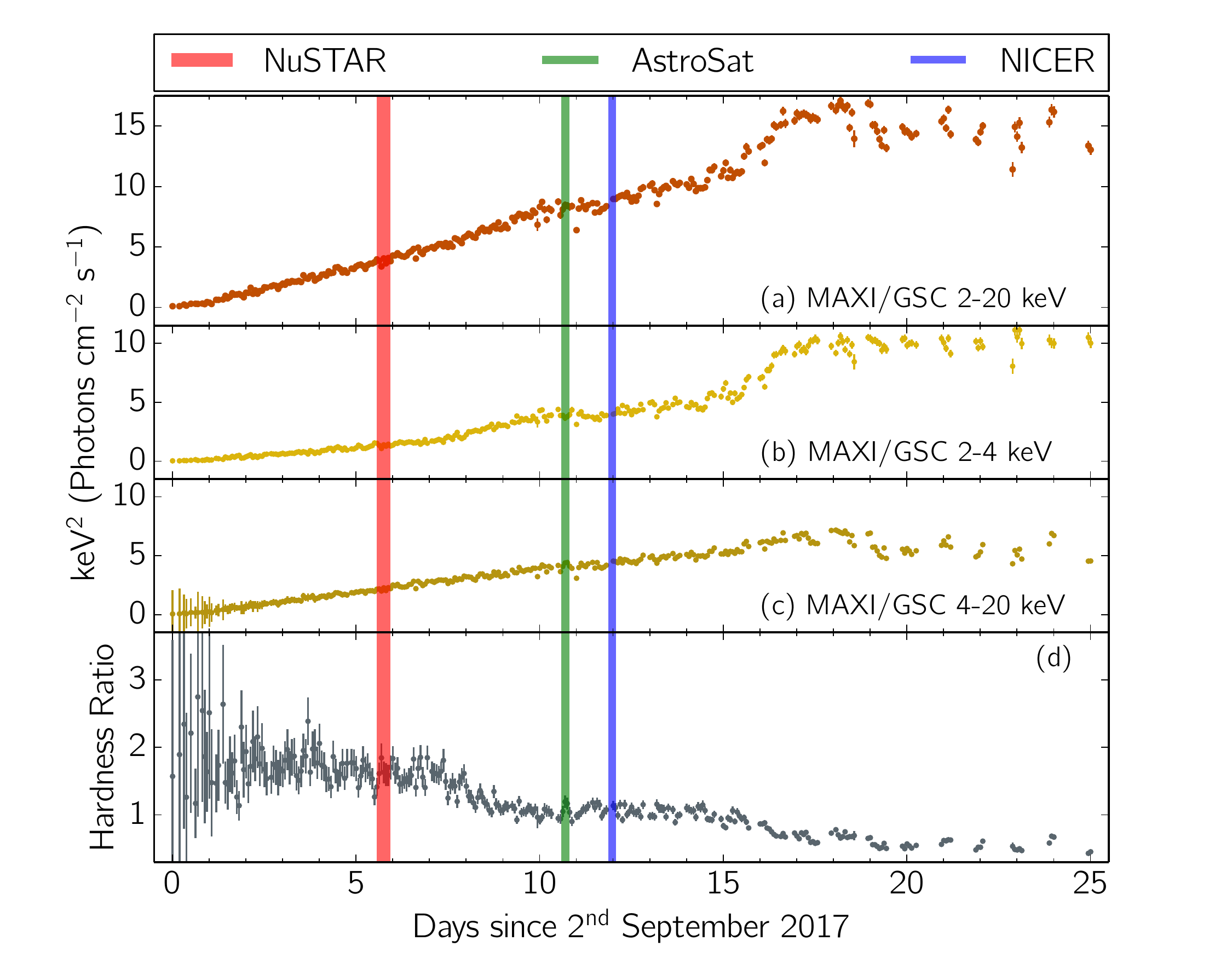}
    \caption{Evolution of \textit{MAXI} intensities and hardness ratio during the initial days of the MAXI J1535--571 outburst. The first (red) vertical band marks the \textit{NuSTAR} observation of the source as reported in \protect\cite{2018ApJ...852L..34X}, the second (green) band marks the \textit{AstroSat} observation primarily used in this paper, and the third (blue) band marks the \textit{NICER} observation as reported in \protect\cite{2018ApJ...860L..28M}. (a) \textit{MAXI}/GSC 2--20 keV orbit-wise lightcurve. (b) \textit{MAXI}/GSC 2--4 keV orbit-wise lightcurve. (c) \textit{MAXI}/GSC 4--20 keV orbit-wise lightcurve. (d) Hardness Ratio is defined as the ratio of the \textit{MAXI}/GSC counts between (4--20) keV and (2--4) keV ranges (See section \ref{obs}). }
    \label{fig:HID}
\end{figure}

The newly discovered black hole X-ray binary MAXI J1535$-$571 was observed as a ToO campaign with \textit{AstroSat}'s co-aligned X-ray instruments - Soft X-ray Telescope (SXT) and Large Area X-ray Proportional Counter (LAXPC). The observations (Obs Id: T01\_191T01\_9000001536) spanned from 2017 September 12 05:32:35 till 2017 September 17 03:42:36. The source remained at a level of $4\sim5$ Crab throughout the campaign. In this work, we focus on the broadband X-ray spectral characterization of MAXI J1535--571 for the first time with \textit{AstroSat}, and analyze the data corresponding to one satellite orbit (orbit 10588; 2017 September 12) which coincides with the onset of the transition to intermediate states of the source (Figure \ref{fig:HID}).

The total source counts detected with \textit{AstroSat}-SXT (hereafter SXT, Energy Band: 1.3--8.0 keV) and \textit{AstroSat}-LAXPC unit-1 (hereafter, LAXPC10 unless specified; Energy Band: 3.0--70 keV) are $\sim 1.50\times 10^{5}$ counts and $\sim 7.07\times 10^{6}$ counts, respectively. Therefore, the analysed data  have reasonably good statistics and hence are enough to serve our purpose of characterising the spectral behaviour of the source during the state transition. Moreover, since we consider the data for only one orbit from the SXT and the LAXPC, our findings are not affected by moderate evolution of source intensity and spectrum during the entire \textit{AstroSat} campaign.

Standard data analysis procedures for individual instruments, as suggested by respective instrument teams, are adopted. The standard data reduction pipelines and tools, disbursed by \textit{AstroSat} Science Support Center (ASSC)\footnote{Processing pipelines, CALDB and response files for all instruments on board \textit{AstroSat} can be downloaded from \url{http://astrosat-ssc.iucaa.in.}}, are utilized to perform the data analysis. Using the \textit{ftool} \texttt{GRPPHA}, channels of each instrument are grouped according to their respective energy resolutions, and a systematic uncertainty of 1.5\% is added to take care of uncertainties in the detector response. Whenever data from both SXT and LAXPC are fitted together in this analysis, we follow the standard method of using a multiplicative constant (denoted by \texttt{`cons'} in Table \ref{tab:models}), to account for differences in flux normalization between the detectors. Following subsections summarize the data analysis methodologies for individual instruments.

\subsection{SXT}
\label{sec:sxt}

SXT\footnote{\url{http://www.tifr.res.in/~astrosat_sxt/index.html}} \citep{2016SPIE.9905E..1ES, 2017JApA...38...29S} on board \textit{AstroSat} is an X-ray telescope operational in the soft energy range of 0.3--8.0 keV. The SXT data, observed in the smaller central window ($10'\times10'$ out of the entire $40'\times40'$ CCD detector), or Fast Window (FW) mode, are processed with \textit{sxtpipeline} v1.4 using the SXT spectral redistribution matrices in CALDB (v20160510). 

\begin{figure}
    \centering
    \includegraphics[width=\columnwidth]{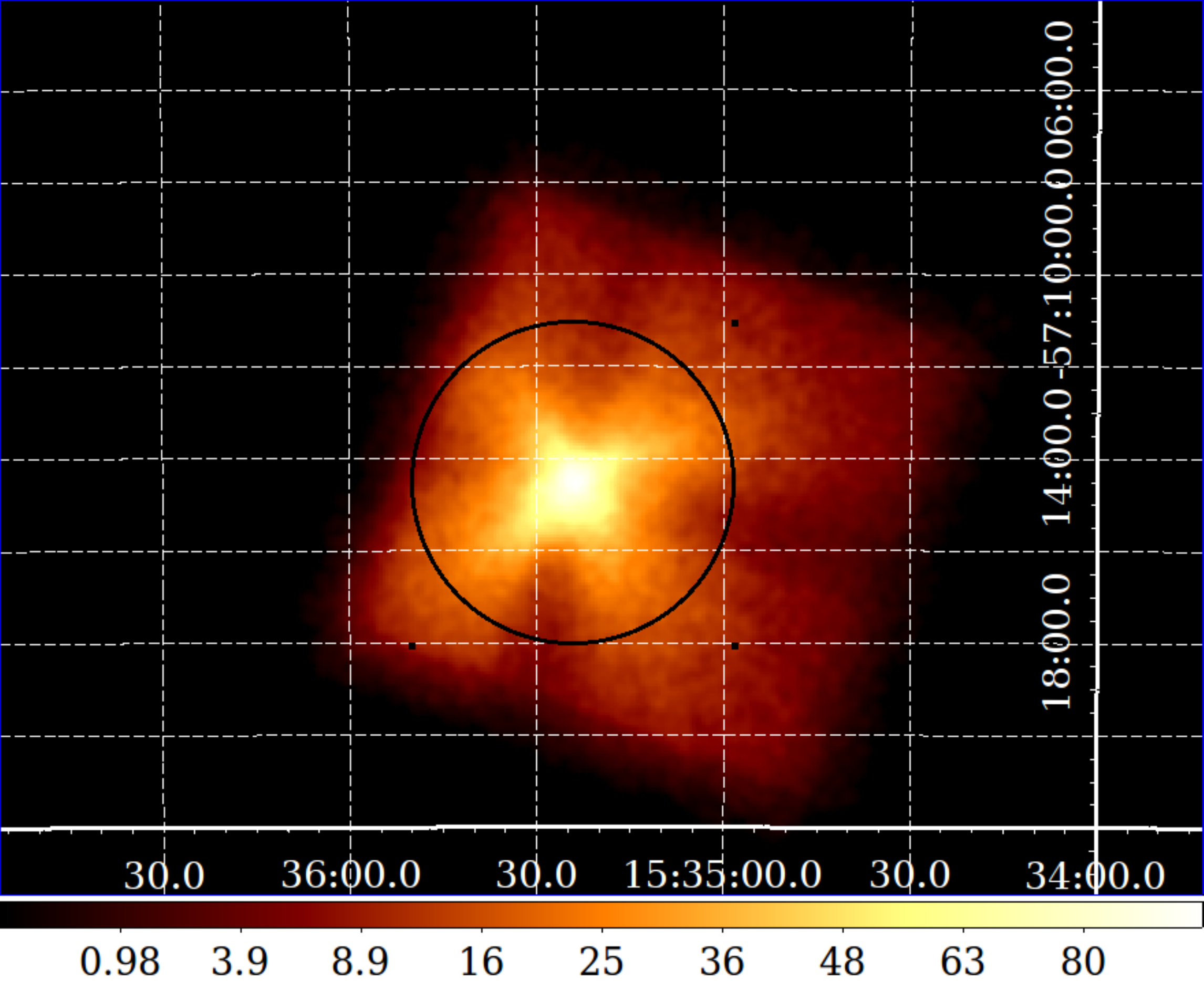}
    \caption{The image displaying the position of source in the SXT detector window and the extraction region (black circle) used for generating the products. The image is smoothed using a Gaussian kernel with 2 pixel radius. The coordinates displayed over the image are obtained from the default coordinate transformation by coordinator tool as a part of the {\tt SXTPIPELINE} (see section~\ref{sec:sxt}).}
    \label{fig:sxtimg}
\end{figure}

We also perform an exercise checking for possible pile-up effects and conclude that the FW mode observations under study are not significantly affected by pile-up, at least not enough to affect our spectral constraints (see Appendix \ref{appendix:pile-up}). The cleaned events file is used to extract image, lightcurves and spectrum utilizing the {\it ftool} {\tt XSELECT}, provided as part of {\tt heasoft-6.24}. A circular region of 3.5 arcmin radius around the source location (RA: {\it 15:35:24.56}, Dec: {\it -57:14:27.23}) is used as the source region (Figure~ \ref{fig:sxtimg}). 

As clearly shown in Figure \ref{fig:sxtimg}, the source is located slightly off-axis in the detector window, and hence the detected source counts are significantly affected by the vignetting effect. The FW mode ancillary response matrix file (ARF) provided by the instrument team is applicable to the source pointed along the center of the FW-window and also for a source extraction region of 5.0 arcmin circular radius. But, we need to use a smaller (3.5 arcmin) radius circular region to keep this region within the source point spread function (PSF) on the SXT CCD, as the FW mode field of view has a size of $10'\times10'$. We, therefore, generate the appropriate ARF, useful for this particular case. This ARF is corrected for the reduced PSF and for vignetting effects due to off-axis pointing and is tested against the Crab data. A co-author (SC), who is a part of the SXT team, is responsible for generating the ARF for this non-standard, off-axis FW mode observation. The deep blank sky background spectrum, provided by the instrument team, is used during spectral modelling. The channels are re-binned to match the intrinsic spectral resolution of SXT\footnote{Page 52 of \textit{AstroSat} handbook v1.11.}. The SXT spectrum for energy band 1.3--8.0 keV is used for combined spectral modelling. The photons below 1.3 keV and above 8.0 keV are ignored to avoid larger systematic errors.

\subsection{LAXPC}
\label{sec:laxpc} 
The LAXPC intrument on board \textit{AstroSat} \citep{2006AdSpR..38.2989A, 2017ApJS..231...10A, 2017JApA...38...30A} consists of three co-aligned nominally identical proportional counter units (LAXPC10, 20 and 30), capable of undertaking X-ray spectral and timing studies in the 3.0--80.0 keV energy range. It has a spectral resolution of $\sim$15\%, and deadtime of 43 $\mu s$, whose effects are corrected for in the light curve and spectrum. All observations were done in the Event Analysis (EA) mode, in which the arrival time of each photon is recorded with an absolute time resolution of 10 $\mu s$. Due to the gain instability issue caused by a gas leakage, the LAXPC30 data are not useful. Note that the data of LAXPC10 are sufficient for a source as bright as MAXI J1535--571, and therefore we do not include data from LAXPC20. In this study, we analyze the data from all the layers of the LAXPC10 detector in the energy range 3.0--70.0 keV. We do not use 70.0--80.0 keV to avoid a larger systematic error. Data are analyzed using the \textit{LaxpcSoft}\footnote{\url{http://www.tifr.res.in/~astrosat_laxpc/LaxpcSoft.html}} software, from which the background and response files are also extracted. LAXPC background is estimated from observation of blank sky where there are no known X-ray sources. We would like to emphasize that, even at the higher energy bands considered (50--70keV), the total (source+background) count rate is consistently higher than the background count rate by a factor of $\sim2$. The count rate obtained during this blanck sky observation is then fitted to latitude and longitude, and this fitted background is subtracted from the observed source counts to get the light curve due to source. Same is done for the spectrum, except that it is averaged over the duration of the observation. The LAXPC channels are logarithmically grouped in the considered energy range.

\section{Spectral Analysis Results}\label{results}

{The spectral fitting and statistical analysis are carried out using the \textsc{xspec} package v-12.9.1q \citep{1996ASPC..101...17A}, distributed as part of {\tt heasoft-6.24} package.}

We adopt several physical and phenomenological models (M1--M7 in Table \ref{tab:models}) to characterize the 1.3--70.0 keV broadband spectra of MAXI J1535--571. All models include the Galactic absorption effect by implementing the \texttt{TBabs} model component with the corresponding abundances and cross sections set according to the \cite{2000ApJ...542..914W} and \cite{1996ApJ...465..487V} photoelectric cross sections. The best fit parameters and the 90\% confidence ranges obtained for these models are listed in Table \ref{tab:bestfit}. We start our analysis by modelling the soft X-ray spectrum of \textit{AstroSat}/SXT with a simple continuum fitting model. We then include the harder X-ray spectrum from LAXPC, and model the resultant broadband continuum spectrum with complicated models accounting for reflection features and the higher energy spectral counterparts. Table \ref{tab:models} shows the progression of models applied to the MAXI J1535--571 data.

\subsection{Without reflection models}\label{without}

First, we fit the 1.3--8.0 keV SXT spectrum with an absorbed multicolour disc blackbody ({\tt diskbb}) plus thermal Comptonization model ({\tt Nthcomp}) (M1). The total setup of model M1 is (Table \ref{tab:models}): \texttt{tbabs*(diskbb+Nthcomp)}. With this fit, we find a $\chi^{2}_{\nu}$ of 1.706 (where $\chi^{2}_{\nu}=\chi^2/\nu$ is the reduced $\chi^{2}$, and $\nu$ is the number of degrees of freedom). Adding a \texttt{Gaussian} component yields a much better $\chi^{2}_{\nu}$ of 1.065 (model M2), indicating the presence of a broad Fe K$\alpha$ emission line, which was found to peak at 6.53 keV ($\sigma=$ 0.53 keV). Note that a broad Fe K$\alpha$ line is not unexpected, because such a line should be the most prominent reflection feature in the 1.3--8.0 keV range, and a reflection spectral component has been found with \textit{NuSTAR} \citep{2018ApJ...852L..34X} and \textit{NICER} \citep{2018ApJ...860L..28M}. Model M2 also gives more constrained values of some parameters, and the \texttt{Nthcomp} best-fit parameter values are close to those found by fitting of \textit{Swift}/XRT spectra (0.6--10.0 kev) of MAXI J1535--571 \citep{StieleKong2018}.

For all our models (Table \ref{tab:models}), we use the thermal Comptonization component \texttt{Nthcomp} (developed by \cite{1996MNRAS.283..193Z} and extended by \cite{1999MNRAS.309..561Z}) to describe the Comptonized component of the continuum spectrum. \texttt{Nthcomp} is a physical model, and hence is better than phenomenological models \citep[e.g., the  powerlaw-based used in ][]{2018MNRAS.480.4443T} to understand the physics of the system.
Particularly, \texttt{Nthcomp} accounts for the low energy rollover around seed photon energies. Typically, these input seed photons can originate from the accretion disc. Hence, we tie the \texttt{Nthcomp} input seed photon temperature to the inner disc temperature of the \texttt{diskbb} component. This best-fit temperature ($\sim 0.3$~keV) obtained from our analysis (Table \ref{tab:bestfit}) is close to the \texttt{diskbb} temperature \citep{2018MNRAS.480.4443T, 2018ApJ...852L..34X} and the \texttt{Nthcomp} seed photon temperature \citep{StieleKong2018} found with \textit{Swift} and \textit{NuSTAR}.

\begin{figure}
	\includegraphics[width=\columnwidth]{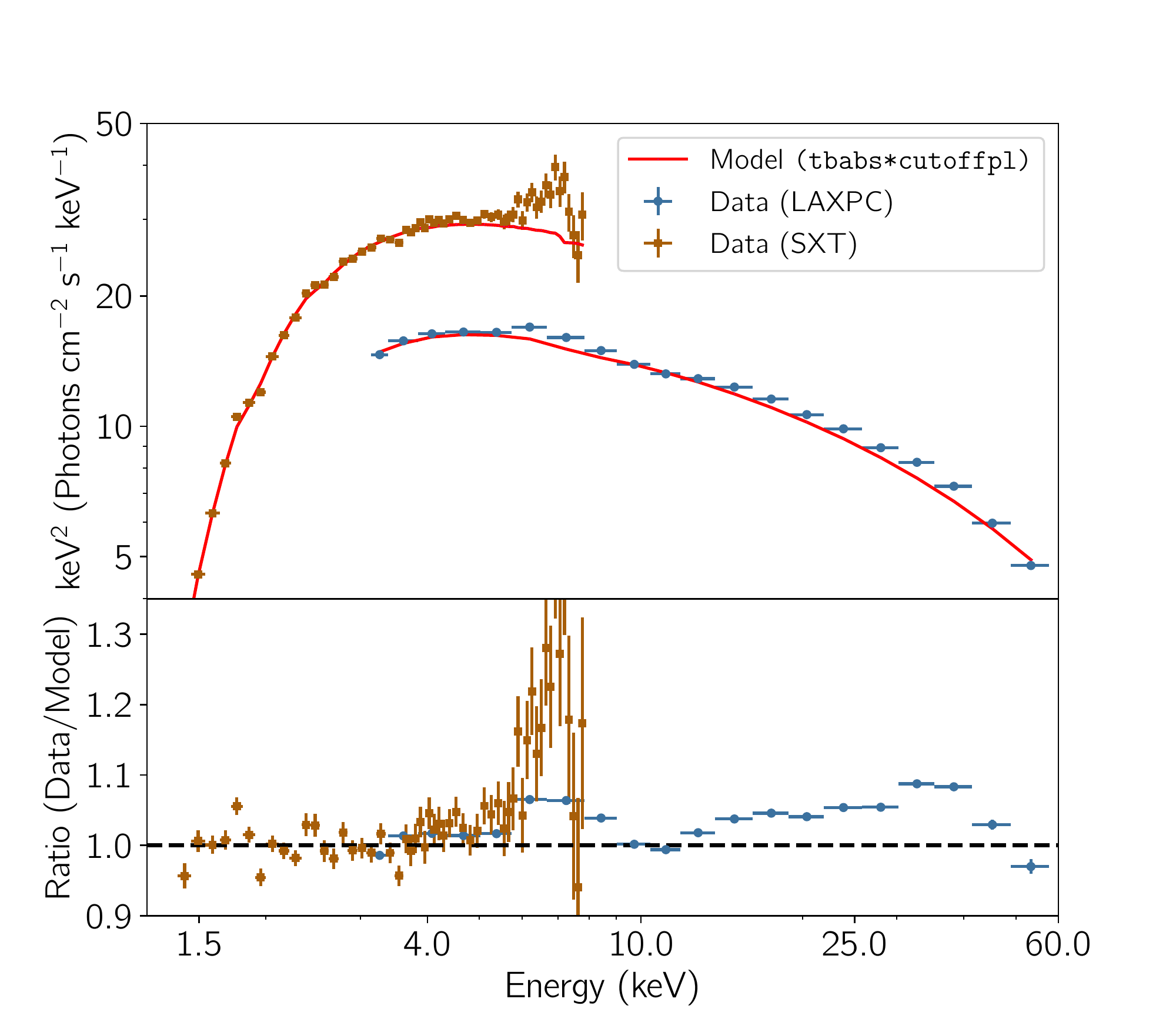}
    \caption{Top panel shows the unfolded \textit{AstroSat} SXT (brown square) and LAXPC (blue circle) spectrum of MAXI J1535--571. The spectrum was fit with a simple absorbed \texttt{cutoffpl} model (solid red line) in the energy ranges: (1.3--4.0) keV, (9.0--12.0) keV and (40.0--60.0) keV. The bottom panel shows the data/model ratio plot, where the reflection features including the broad Fe K$\alpha$ emission line at $\sim 6.5$~keV and the Compton hump at $\sim 30$~keV can be clearly seen (see section~\ref{without}).}
    \label{fig:ratio}
\end{figure}

As the next step, we include the hard X-ray spectrum in our analysis. We start with the combined SXT and LAXPC10 spectrum, and first try to check if the reflection features, found with \textit{NuSTAR} \citep{2018ApJ...852L..34X} and \textit{NICER} \citep{2018ApJ...860L..28M}, can be seen. For this purpose, we rebin the channels of both the instrument according to their respective energy resolutions, and fit the spectrum with a simple absorbed cutoff-powerlaw model (\texttt{tbabs*cutoffpl} in \textsc{xspec}), only in the energy ranges: 1.3--4.0 keV, 9.0--12.0 keV and 40.0--60.0 keV. These are the energy ranges where the reflection features from the accretion disc are not prominently expected to be seen. Data/model ratio plot (Figure \ref{fig:ratio}) corresponding to this fit exhibits a large residual which includes the signatures of main reflection components: a broad Fe K$\alpha$ line at $\sim 6.5$ keV and a Compton hump peaking at $\sim 30$ keV. In the residual, we also see a peak at $\sim 33$ keV that could be because of the Xenon K emission feature. This feature is not observed in the \textit{AstroSat}-CZTI spectra (not included here), indicating that this is of instrumental origin. In order to account for this, we include an ad hoc \texttt{Gaussian} component in all fits that include the LAXPC10 spectrum (denoted by \texttt{`G'} in Table \ref{tab:models}). With fits, its line energy and width constrain between 31--35 keV and $\sigma \sim$ 1.0--5.0 keV respectively. The obtained values are then frozen while fitting for the other components. The inclusion of this frozen Gaussian component has minimal and cosmetic effect on the other model parameters.

\begin{figure}
	\includegraphics[width=\columnwidth]{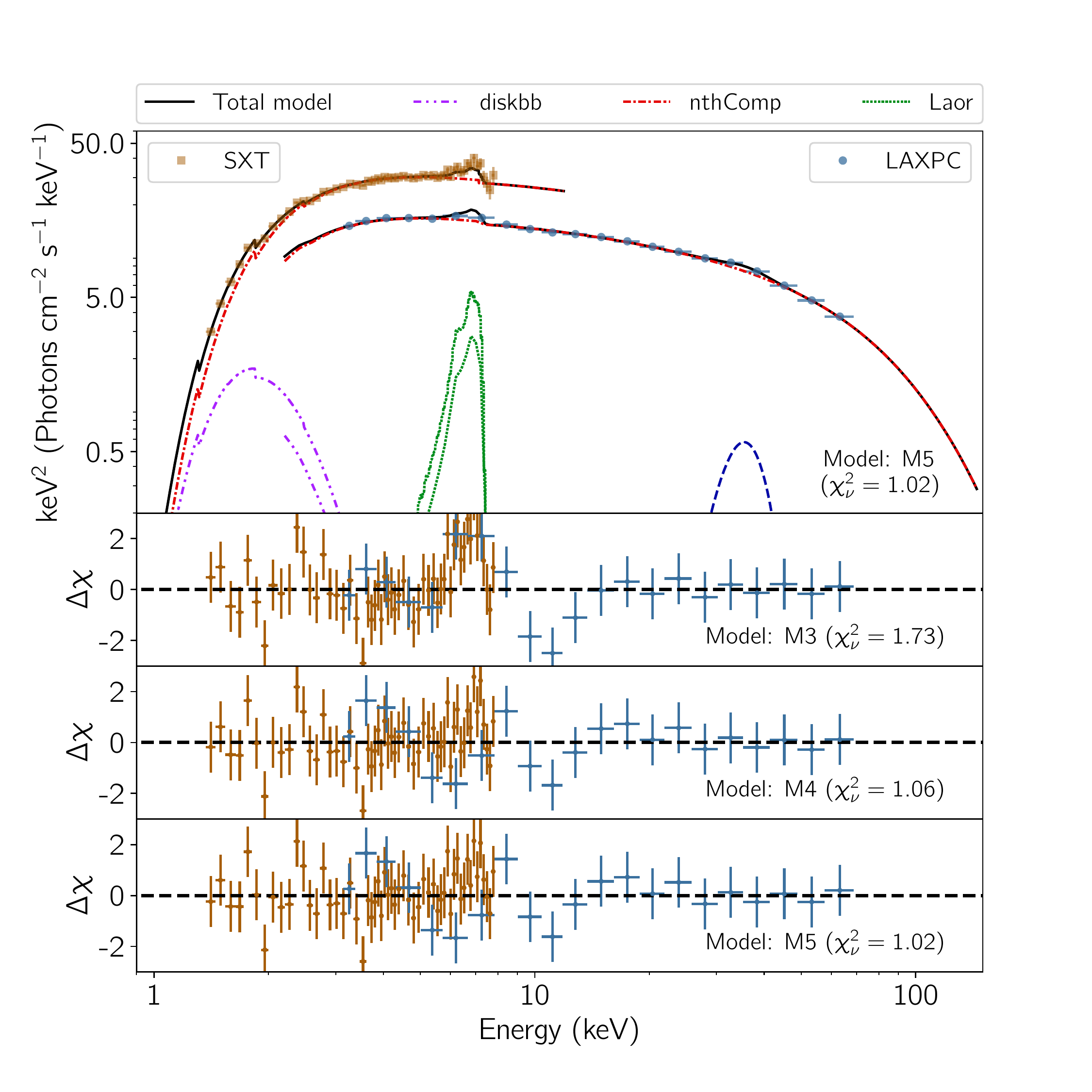}
    \caption{Top panel represents the extrapolated model (M5) of the continuum fit (solid black lines), the binned spectral data of SXT (brown square) and LAXPC (blue circle) instruments for MAXI J1535--571, and the individual components of the model. The components are, \texttt{diskbb} (magenta dashdotdotdash), \texttt{Nthcomp} (red dashdotdash), \texttt{Laor} (green dot) and \texttt{Gaussian (`G')} (dark blue dash). The three bottom panels show the evolution of residuals as the model progresses from M3 to M5 (Table \ref{tab:models}). Clear sign of the Fe K$\alpha$ emission feature is seen in M3, which fades away as we account for that in the subsequent models with \texttt{Gaussian} (M4) and \texttt{Laor} (M5) components (see section~\ref{without}).}
    \label{fig:M3-M5}
\end{figure}

We then perform a combined fit of the SXT and LAXPC spectra with an absorbed disc  blackbody and thermal Comptonization model - M3 (\texttt{cons*tbabs*(diskbb+Nthcomp+G)} in \textsc{xspec}). This fit yielded a $\chi^{2}_{\nu}$ of 1.726. For all the fits that involve multiple instruments, a normalization constant (\texttt{cons} in Table \ref{tab:models}) is included to account for the differences in the flux calibration between SXT and LAXPC. Despite the inclusion of a cross normalization constant, this model (M3) fails to adequately describe the spectrum in the 5.0--11.0 keV range (Figure \ref{fig:M3-M5}, panel 2). We then include a \texttt{Gaussian} component and fit for its central line energy and line width constrained between 6.2--6.8 keV and 0.05--1.0 keV respectively. Inclusion of this \texttt{Gaussian} component is seen to considerably improve the fit (Table \ref{tab:models}), yielding a $\chi^{2}_{\nu}$ of 1.063 (M4). The corresponding best-fit parameters are given in  Table \ref{tab:bestfit}. In order to probe this Fe K$\alpha$ emission feature (also seen in Figure \ref{fig:ratio}), we replace the \texttt{Gaussian} component with the \texttt{Laor} component, which includes general relativistic effects of a maximally spinning black hole \citep{1991ApJ...376...90L}. 

 We find that the data cannot simultaneously constrain all the free parameters well. So, we freeze the power law index ($\alpha$) of the emissivity profile ($\epsilon(r)\propto r^{-\alpha}$) to the canonical standard accretion disc value of 3 \citep{1989MNRAS.238..729F}. We also freeze the outer disc radius ($R_\mathrm{out}$) at 400$R_\mathrm{g}$, and fit for the line's central peak, inner disc radius and the disc's inclination (see Table \ref{tab:bestfit}). Adding a \texttt{Laor} component in place of a \texttt{Gaussian} improves the fit modestly (Table \ref{tab:models}), with the $\chi^{2}_{\nu}$ decreasing from 1.063 (M4) to 1.016 (M5). With this model, we constrain the inner disc truncation radius and disc inclination to $5.1_{-2.8}^{+9.2} R_\mathrm{g}$ and $45_{-9}^{+34^\circ}$ respectively. The top panel of Figure \ref{fig:M3-M5} shows the best fitted model (M5), and the bottom three panels show the data/model ratio plots for models M3, M4 and M5. Fe K$\alpha$ emission feature can be clearly seen in the ratio plot of M3, and it fades away as the model evolves to include a \texttt{Laor} component (M5) to accommodate for the Fe K$\alpha$ emission complex.

\begin{table*}
 \caption{Progression of various models used for fitting the \textit{AstroSat} spectra of MAXI J1535--571, and the corresponding statistics (see section~\ref{results}).}
 \label{tab:models}
 \hspace{-0.7cm}
 \begin{tabular}{lllllll}
   
   \hline
   \hline
  Model & $\chi^{2}$ & $\nu$ & $\chi^{2}_{\nu}$ & $\Delta \chi^{2}/\Delta \nu$ & \textit{f-test} & Instrument(s)\\  &  &  &  &  & probability & \\
  \hline
  
 M1:  \texttt{tbabs*(diskbb+Nthcomp)} & 85.28 & 50 & 1.706 & - & - & SXT\\

 M2:  \texttt{tbabs*(diskbb+Nthcomp+Gaussian)} & 50.07 & 47 & 1.065 & 11.74 & - & SXT\\

 M3:  \texttt{cons*tbabs*(diskbb+Nthcomp+G)} & 120.81 & 70 & 1.726 & - & - & SXT+LAXPC\\

 M4:  \texttt{cons*tbabs*(diskbb+Nthcomp+Gaussian+G)} & 71.20 & 67 & 1.063 & 16.54 & - & SXT+LAXPC\\
 
 M5: \texttt{cons*tbabs*(diskbb+Nthcomp+Laor+G)} & 67.05 & 66 & 1.016 & 4.15 & 0.085 & SXT+LAXPC\\
 
  \hline

 M6: \texttt{cons*tbabs*(Kerrbb+Nthcomp+relxilllpCp+G)} & 78.40 & 61 & 1.285 & - & - & SXT+LAXPC\\
  
 M7: \texttt{cons*tbabs*(Kerrbb+Nthcomp+relxilllpCp+xillverCp+G)} & 70.15 & 60 & 1.169 & 8.25 & 0.010 & SXT+LAXPC\\
  
   \hline
 \end{tabular}
\begin{flushleft}
 \textbf{Note:} `\texttt{G}' denotes the addition of a fixed \texttt{Gaussian} component to all the LAXPC10 spectra in order to account for the Xenon K emission feature at $\sim 33$ keV. $\Delta \chi^{2}/\Delta \nu$ and \textit{f-test} probability values are calculated with respect to the model listed in the previous line, and only for the same set of instruments. \textit{F-test} probability values are not estimated for models involving an addition of a Gaussian component \citep{2002ApJ...571..545P}.
\end{flushleft}
\end{table*}

\subsection{With reflection models}\label{with}

As mentioned earlier (section~\ref{Introduction}), a reflection spectral component of MAXI J1535--571 has been detected with \textit{NuSTAR} \citep{2018ApJ...852L..34X} and \textit{NICER} \citep{2018ApJ...860L..28M}. In section~\ref{without}, we took care only of the Fe K$\alpha$ emission line profile of the reflection spectrum, using the \texttt{Gaussian} or \texttt{Laor} component. In this section, we attempt to characterize the full reflection component (Fe K$\alpha$ line, Compton hump, absorption edge, etc.) using the combined 1.3--70.0 keV SXT+LAXPC spectrum with a more physically realistic \textsc{relxill}\footnote{\url{http://www.sternwarte.uni-erlangen.de/research/relxill}} family of relativistic reflection models \citep{2014MNRAS.444L.100D, 2014ApJ...782...76G}. 

Here, the non-reflection part of the continuum spectrum is fitted with a multi-temperature blackbody ({\texttt{Kerrbb}}) and a thermal Comptonization model (\texttt{Nthcomp}), with the interstellar absorption accounted for. The input seed photon temperature ($kT_\mathrm{bb}$) in the \texttt{Nthcomp} component is frozen to a value of 0.26 keV, as obtained from M5. We start modelling the reflection spectrum using the \texttt{relxilllpCp} component of the model, as this component requires lesser number of parameters, takes care of the smearing of reflection features due to relativistic effects in the vicinity of the black hole and also parameterizes the emissivity profile in physical units. The shape of the illuminating continuum that the \texttt{relxilllpCp} model assumes is that of the thermally Comptonized spectrum, and the Comptonizing source assumes a lamp-post geometry in which the condensed corona is located at the rotation axis above the black hole, at a height \textit{h}. Due to the assumption of a thermally Comptonized illuminating spectrum, relevant parameters from the \texttt{Nthcomp} component like the spectral index and electron temperature ($kT_\mathrm{e}$) are tied with the ones in \texttt{relxilllpCp} component. In this model, the reflection fraction is self-consistently determined from ray-tracing calculations based on the values estimated for the inner disc radius ($R_\mathrm{in}$), the black hole spin parameter ($a_{\star}$) and the height of the coronal lamp post ($h$). To describe the disc spectrum, we use the \texttt{Kerrbb} model \citep{2005ApJS..157..335L}, which assumes a multi-temperature blackbody spectrum from a thin steady state general relativistic accretion disc around a Kerr black hole. While modelling for the disc spectrum using \texttt{Kerrbb}, we consider the effects of self-irradiation, and assume the torque at the inner boundary of the disc to be zero \citep{1973A&A....24..337S, 1973blho.conf..343N}.
 \cite{1995ApJ...445..780S} prescribed a value of the colour factor $f_\mathrm{col}$ = 1.7$\sim$2.0 for very high accretion rates close to the Eddington Luminosity, where $f_\mathrm{col}$ is defined as the ratio of the colour temperature to the effective temperature ($T_\mathrm{col}/T_\mathrm{eff}$). Otherwise, for a few times $0.1 L_{\rm Edd}$, they prescribed an $f_\mathrm{col}$ of 1.7, which we adopt in our case. Thus, the total set up for model M6 is (Table~\ref{tab:models}): \texttt{cons*tbabs*(Kerrbb+Nthcomp+relxilllpCp+G)}. 

\begin{table*}
 \caption{Best fit parameters of the \textit{AstroSat} spectra of MAXI J1535--571 and the corresponding 90\% confidence ranges obtained for different models M1--M7 given in Table~\ref{tab:models} (see section~\ref{results}).}
 \label{tab:bestfit}
\renewcommand\tabcolsep{1pt}
 \begin{tabular}{lllllllllll}
   
   \hline
   \hline
  Spectral & Parameters & M1 & M2 & M3 & M4 & M5 & M6 & M7\\
  components & & & & & & & & & \\
  \hline
  
  \verb'TBabs' & $N_\mathrm{H}$\footnotemark[1] $(\times$ 10$^{22}$cm$^{-2}$) & $2.41_{-0.16}^{+0.14}$ & $2.66_{-0.17}^{+0.17}$ & $3.14_{-0.16}^{+0.15}$ & $2.92_{-0.25}^{+0.17}$ & $2.87_{-0.23}^{+0.15}$ & $2.85_{-0.14}^{+0.13}$ & $2.79_{-0.09}^{+0.11}$\\

  &&&&&&&&&\\

  \verb'diskbb' & $T_\mathrm{in}$\footnotemark[2] (keV) & $0.22_{-0.05}^{+0.24}$ & $0.34_{-0.07}^{+0.02}$ & $0.25_{-0.01}^{+0.01}$ & $0.26_{-0.03}^{+0.02}$ & $0.26_{-0.04}^{+0.02}$ &  ... & ... \\
[1ex]   
   & norm\footnotemark[3] $(\times 10^5)$ & $<1.25$ & $2.8_{-1.6}^{+5.0}$ & $43_{-16}^{+29}$ & $19_{-11}^{+18}$ & $15_{-9}^{+14}$ & ... & ... \\
  
  &&&&&&&&&\\
  
  \verb'Nthcomp' & $\Gamma_\mathrm{Nthcomp}$\footnotemark[4] & $2.17_{-0.14}^{+0.01}$ & $2.04_{-0.02}^{+0.17}$ & $2.30_{-0.01}^{+0.02}$ & $2.28_{-0.02}^{+0.01}$ & $2.27_{-0.02}^{+0.01}$ & $2.25_{-0.01}^{+0.01}$ & $2.25_{-0.01}^{+0.01}$ \\
  [1ex]
   &$kT_\mathrm{e}$\footnotemark[5] (keV) & $>85.54$ & $2.6_{-1.1}^{+5.6}$ & $21.9_{-1.3}^{+1.9}$ & $21.2_{-1.3}^{+1.5}$ & $20.9_{-1.3}^{+1.6}$ & $20.02_{-0.60}^{+0.63}$ & $19.70_{-0.50}^{+0.61}$ \\
   [1ex]
  & norm\footnotemark[6] & $41_{-17}^{+2}$ & $32_{-2}^{+11}$ & $49.9_{-1.7}^{+2.2}$ & $46.9_{-2.3}^{+2.2}$ & $46.1_{-2.4}^{+2.5}$ & $39.3_{-1.4}^{+1.7}$ & $37.8_{-1.9}^{+0.7}$  \\

  &&&&&&&&&&\\
  
  \verb'Gaussian' & $E_\mathrm{Gau}$\footnotemark[7] (keV) & ... & $6.53_{-0.13}^{+0.16}$ & ... & $6.53_{-0.12}^{+0.15}$ &  ... & ...  & ... \\
  [1ex]
  & $\sigma$\footnotemark[8] (keV) & ... & $0.53_{-0.18}^{+0.16}$ & ... & $0.50_{-0.17}^{+0.26}$ & ... & ...  & ... \\
  [1ex]
  & norm\footnotemark[9] &...  & $0.24_{-0.11}^{+0.07}$ & ... & $0.13_{-0.03}^{+0.04}$ & ... &... & ... \\

  &&&&&&&&&&\\
  
  \verb'Laor' & $E_\mathrm{Laor}$\footnotemark[10] (keV) & ... & ... &  ...& ... & $6.53_{-0.23}^{+0.15}$ & ... & ...\\
  [1ex]
  & R$_\mathrm{in}$\footnotemark[11] ($R_\mathrm{g}$) & ... & ... &  ...& ... & $5.1_{-2.8}^{+9.2}$ & ... & ... \\
  [1ex]
  & \textit{i}\footnotemark[12] ($^\circ$) & ... & ... &  ...& ... & $45_{-9}^{+34}$ & ... & ...\\
  [1ex]
  & norm\footnotemark[13] & ... & ... &  ...& ... & $0.14_{-0.03}^{+0.04}$ & ... & ...\\
   
  &&&&&&&&&&\\
  
  \verb'Kerrbb' & $M_\mathrm{BH}$\footnotemark[14] ($M_{\odot}$) & ... & ... & ... & ... & ...  & $10.3_{-1.7}^{+0.6}$ & $10.39_{-0.62}^{+0.61}$\\
  [1ex]
  & \.{M}\footnotemark[15] $(\times 10^{17}$ g s$^{-1})$  & ... & ... & ... & ... & ... & $0.25_{-0.01}^{+0.08}$ & $0.31_{-0.04}^{+0.03}$\\
  [1ex]
  & $D_\mathrm{BH}$\footnotemark[16] (kpc)  &...  & ... & ... &...  & ...  & $5.3_{-1.0}^{+0.9}$ & $5.4_{-1.1}^{+1.8}$\\		
  [1ex]
  & norm\footnotemark[17]  & ... & ... & ... & ... & ... & $139_{-38}^{+74}$ & $151_{-68}^{+82}$\\
  
  &&&&&&&&&&\\
  
   \verb'relxilllpCp' & \textit{h}\footnotemark[18] ($R_\mathrm{g}$) & ... & ... & ... & ... & ...& $7.3_{-1.1}^{+1.3}$ & $9.34_{-0.27}^{+0.25}$\\
   [1ex]
   & $a_{\star}$\footnotemark[19] (cJ/GM$^{2}$) & ... & ... & ... & ... & ... &  $0.67_{-0.12}^{+0.11}$ & $0.67_{-0.04}^{+0.16}$\\
   [1ex]
   & \textit{i}\footnotemark[20] ($^\circ$) & ... & ... & ... & ... & ... & $40.9_{-5.5}^{+5.7}$ & $79.9_{-4.0}^{+4.2}$\\
   [1ex]
   & $R_\mathrm{in}$\footnotemark[21] ($R_\mathrm{ISCO}$) & ... & ... & ... & ... & ... & <1.61 & <1.23\\
   [1ex]
   & $\log{\xi}$\footnotemark[22] (log[erg cm s$^{-1}$]) & ...  & ... & ... & ... & ... & $3.59_{-0.17}^{+0.32}$ & $3.74_{-0.25}^{+0.31}$\\
   [1ex]
   & norm\footnotemark[23] & ... & ... & ... & ... & ... &  $0.12_{-0.02}^{+0.02}$ & $0.08_{-0.04}^{+0.06}$\\
     
  &&&&&&&&&&\\
  
   \verb'xillverCp'  & norm\footnotemark[24] & ... & ...  & ... & ... & ... & ... & $0.15_{-0.08}^{+0.06}$\\
   
      \hline
   Unabsorbed flux & 1.3--8.0 keV & 10.64 & 11.58 & 13.69 & 12.63 & 12.43 & 11.9  & 13.56 \\
     [1ex]
   (10$^{-8}$erg cm$^{-2}$ s$^{-1}$) & 3.0--70.0 keV & 13.78 & 7.59 & 11.73 & 11.74 & 11.72 & 11.64  & 13.14 \\
    [1ex]
     & 2.0--10.0 keV & 8.89 & 8.96 & 9.53 & 9.36 & 9.31  & 9.20 & 10.56 \\
   \hline
     & $C_\mathrm{LAXPC}$\footnotemark[25] & ... & ... & $0.81_{-0.01}^{+0.01}$ & $0.80_{-0.01}^{+0.01}$& $0.80_{-0.01}^{+0.01}$ & $0.80_{-0.01}^{+0.01}$ & $0.80_{-0.01}^{+0.01}$\\
     [1ex]

        \hline   
 \end{tabular}
 \begin{flushleft}
 \textbf{Note:}
 $^1$ Hydrogen column density; $^2$ Temperature at the inner disc radius; $^4$ Asymptotic power-law photon index; $^5$ Electron temperature determining the high energy rollover; $^{7}$ Gaussian line energy; $^{8}$ Gaussian line width; $^{10}$ Line energy; $^{11}$ Inner radius of the accretion disc (in units of gravitational radii $R_\mathrm{g}$); $^{12}$ Disc inclination; $^{14}$ Mass of the black hole; $^{15}$ Effective mass accretion rate; $^{16}$ Distance of the black hole from the observer; $^{18}$ Height of the comptonizing source above the black hole; $^{19}$ Dimensionless spin parameter of the black hole ($a_{\star}$ = $cJ/GM^2$, where $J$ is the angular momentum of the black hole); $^{20}$ Inclination of the inner disk; $^{21}$ Inner-radius of the accretion disc (in units of $R_\mathrm{ISCO}$); $^{22}$ $\log$ of the ionization parameter ($\xi$) of the accretion disk, where $\xi=L/nR^{2}$, with \textit{L} as the ionizing luminosity, \textit{n} as the gas density and \textit{R} as the distance to the ionizing source; $^{3,6,9,13,17,23,24}$ Normalization parameter of the corresponding spectral component; $^{25}$ The flux normalization constant $C_\mathrm{LAXPC}$ for LAXPC (denoted by \texttt{`cons'} in Table \ref{tab:models}) is estimated with respect to the SXT flux.
\end{flushleft}
\end{table*}

\begin{figure}
	\includegraphics[width=\columnwidth]{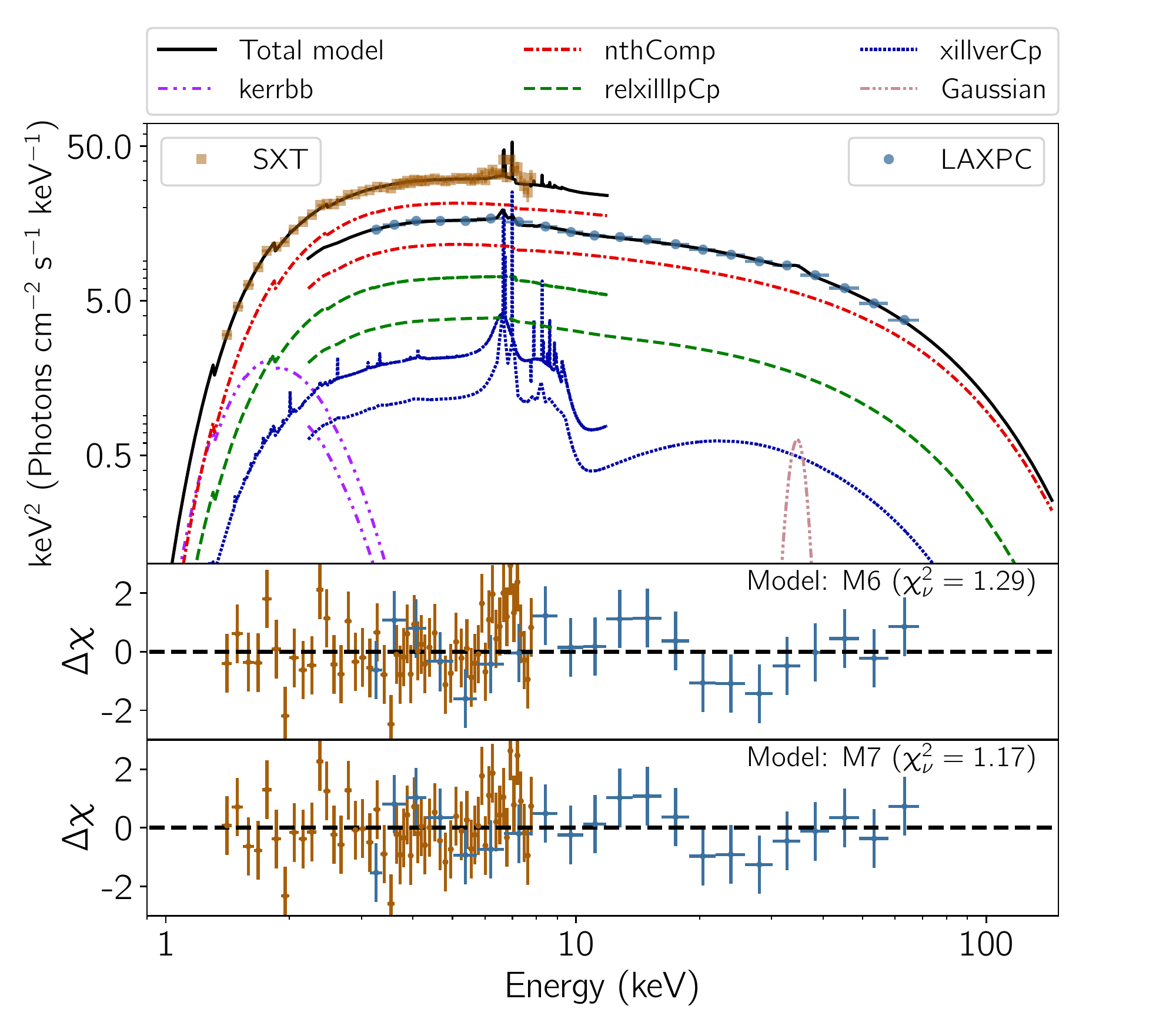}
    \caption{Top panel represents the extrapolated model (M7) of the continuum fit (solid black lines), the binned spectral data of SXT (brown square) and LAXPC (blue circle) instruments for MAXI J1535--571, and the individual components of the model. The components are, \texttt{Kerrbb} (magenta dashdotdotdash, \texttt{Nthcomp} (red dashdotdash), \texttt{relxilllpCp} (green dash), \texttt{xillverCp} (dark blue dot) and \texttt{Gaussian (`G')} (peach dashdotdotdotdash). The two bottom panels show the residuals for the models M6 and M7, the latter includes a distant unblurred reflection component - \texttt{xillverCp}. It can be seen that the addition of the \texttt{xillverCp} component in M7 to a blurred reflection component (as in M7) improves the fit in the Fe K$\alpha$ emission line region (see section~\ref{with}).}
    \label{fig:reflection}
\end{figure}

With this set up, we simultaneously fit for the value of spin $a_{\star}$, inner disc truncation radius $R_\mathrm{in}$, inclination of the disc and its ionization parameter log(${\xi}$), by freezing the outer edge of accretion disc $R_\mathrm{out}$ at 400$R_\mathrm{g}$. $R_\mathrm{out}$ mentioned here, is much smaller than the actual radius of the disc rim, and corresponds only to the radius of the accretion disk, which is relevant for X-ray reflection. Since the \textsc{relxill} family of models assume reflection from a slab, the contribution to reflection from large radii is insignificant, and the precise value of $R_\mathrm{out}$ is thus not important here. We constrain $a_{\star}$ to $0.67_{-0.12}^{+0.11}$, $R_\mathrm{in}$ to <1.61 $R_\mathrm{ISCO}$, inclination to ${40.9_{-5.5}^{+5.7}}$~degree and log($\xi$) to $3.59_{-0.17}^{+0.32}$ log (erg cm s$^{-1}$), where the ionization parameter $\xi=L/nR^{2}$, with \textit{L} as the ionizing luminosity, \textit{n} as the gas density and \textit{R} as the distance to the ionizing source. The same parameters in the reflection component and the \texttt{Kerrbb} component (spin $a_\star$, disc inclination $i$) are tied with each other, and the \texttt{Kerrbb} parameters, i.e., the black hole mass, its distance and the effective mass accretion rate are estimated to be $10.3_{-1.7}^{+0.6}~M_{\odot}$, $5.3_{-1.0}^{+0.9}$ kpc, and $0.25_{-0.01}^{+0.08}\times10^{17}$ g s$^{-1}$ respectively. The model M6, which includes a relativistic reflection component, yields a satisfactory fit to the spectra with a $\chi^{2}_{\nu}$ of 1.285. While this $\chi^{2}_{\nu}$ value is somewhat higher than that for the model M5 (see Table~\ref{tab:models}), the merit of M6 is it, unlike M5, includes the full reflection component. However, the residual in the 6.0--9.0 keV region (middle panel of Figure \ref{fig:reflection}) still indicates the presence of features unaccounted for by the model M6.

Therefore, in addition to the smeared reflection component from the vicinity of the black hole, we also include a possible contribution from reprocessing by distant material using an unblurred reflection component -- \texttt{xillverCp} \citep{2010ApJ...718..695G}. Thus, the total set up for model M7 is (Table~\ref{tab:models}): \texttt{cons*tbabs*(Kerrbb+Nthcomp+relxilllpCp+ xillverCp+G)}. We first assume the unblurred distant reprocessing material to be neutral by fixing $\log(\xi)$ = 0, as neutral narrow Fe K$\alpha$ lines have earlier been found in bright Galactic binaries \citep{2015ApJ...808....9P}. However, this assumption is seen to deteriorate the fit, indicating that the distant reprocessing material is ionized. This is supported by the following. From the fitting with the model M5, we find that the core of the iron line peaks at 6.53 keV (Table~\ref{tab:bestfit}). If this core represents the narrow line originating from the distant material, then its energy, which is higher than a neutral line energy of $\sim 6.4$ keV, indicates the ionization of the distant material.
Allowing ionization parameter of the \texttt{xillverCp} component to vary freely yields a much better fit, and the value of log($\xi$) is estimated to be very close to what is obtained for the \texttt{relxilllpCp} component, within the error bars. So, we perform the fit of M7, with all the \texttt{xillverCp} parameters, except the normalization, tied to the corresponding \texttt{relxilllpCp} component parameters. Both the M6 and M7 models are initially fitted with the iron abundance ($A_\mathrm{Fe}$) as a free parameter. The fits do not yield a constrained value of the parameter, rather is seen to be pegged at value of 5. So for the reflection models, we freeze the value of $A_\mathrm{Fe}$ to 5 times the solar abundance.

The blurred and the unblurred reflection components have a parameter for the reflection fraction, which allows for a thermal Comptonization continuum. As we include an external \texttt{Nthcomp} component accounting for that, we freeze this parameter ($R_{f}$) to --1.  For these fits (M6 and M7), we assume a canonical emissivity profile -- $\epsilon(r)\propto r^{-q}$, where the emissivity index, $q = 3$ \citep{1989MNRAS.238..729F}. As mentioned earlier, the values for the ionization parameter, iron abundance and the input continuum in \texttt{xillverCp} are linked with those of the broad reflection component, as we find no empirical need to decouple those components. It could be seen from the residuals of the fit (bottom panel of Figure \ref{fig:reflection}) that, model M7 yields a better fit to the data with a $\chi^{2}_{\nu}$ = 70.15/60 = 1.17, than M6. As Table~\ref{tab:bestfit} shows, the best-fit values for M7 are also found to be consistent within the error bars with those of the previously fitted model M6, except for the higher values of the height of coronal lamp-post ($9.34_{-0.27}^{+0.25}~R_\mathrm{g}$) and the inclination angle (${79.9_{-4.0}^{+4.2}}$~degree).

Additionally, we perform a couple of other exercises by modifying M7. The aim of these exercises is to verify the suitability of using a continuum model like \texttt{Kerrbb}, for black holes in their intermediate states. First, we let the spectral hardening factor of \texttt{Kerrbb} component to be a free parameter. This leads its parameter to a fit value to $1.73_{-0.21}^{+0.17}$, as opposed to the earlier frozen value of 1.7. The best fit values of other key parameters including the mass, accretion rate and spin parameter of the black hole remain within the earlier estimated confidence ranges. We also note here that, \cite{1995ApJ...445..780S} had demonstrated a weak relationship between the spectral hardening factor and, the mass of the black hole, its spin parameter and the accretion rate. So, it is not surprising to see no considerable changes to these parameters in our exercise, with a free spectral hardening factor. As the second exercise, we decouple the spin and disc inclination parameters of \texttt{Kerrbb} from those of the \textsc{relxill} component. Although their values from the reflection component remain relatively the same, the corresponding parameters in the \texttt{Kerrbb} component are seen to get pegged at their upper limits, and the data could not constrain the error ranges either. This exercise does yield a satisfactory fit, but the $\chi^2_{\nu}$ is found to increase slightly to 1.183 as opposed to 1.169 earlier.

\section{Discussion and conclusions}\label{discussion}

In this paper, we report the results of a broadband spectral characterization of the transient BHXB MAXI J1535--571, as observed with \textit{AstroSat} in 2017 during its hard to intermediate state transition. Our analysis of the source includes data from the SXT (1.3--8.0 keV) and the LAXPC (3.0--70.0 keV) on board \textit{AstroSat}. In this section, we discuss implications of our results, which shows the potential of \textit{AstroSat} to unveil the broadband spectral characteristics of X-ray binaries. We begin our analysis by fitting the observed X-ray spectrum with a simple continuum model consisting of an absorbed multi-colour blackbody from the disc (\texttt{diskbb}) plus a thermal Comptonization component (\texttt{Nthcomp}) from a corona. Note that, while \texttt{diskbb} and \texttt{Nthcomp} are ideal to describe the spectral components from the disc and the corona of BHXBs, the current paper is explicitly using these two physical model components together for MAXI J1535--571 continuum X-ray spectrum for the first time to the best of our knowledge. Here we note that, for all the models we use, the best-fit hydrogen absorption column density value comes out to be in the range of $\sim (2.4-3.1)\times10^{22}$~cm$^{-2}$ (Table~\ref{tab:bestfit}), which is not very different from the values measured with \textit{MAXI}, \textit{Swift}, \textit{XMM-Newton} and \textit{NICER} \citep{2017ATel10729....1N, 2018MNRAS.480.4443T, StieleKong2018, 2018ApJ...860L..28M}. The best-fit \texttt{diskbb} temperature $T_{\rm in}$, for all our models having the \texttt{diskbb} component, is in the range of $\sim 0.2-0.3$~keV. While this is slightly smaller than the \textit{NuSTAR} value of $\sim 0.4$~keV \citep{2018ApJ...852L..34X}, this paper mentions that the \textit{NuSTAR} value is somewhat larger than the values found for some well-known BHXBs in the hard state. We get confidence in our best-fit $T_{\rm in}$ values also from the following. We tie $T_{\rm in}$ with the \texttt{Nthcomp} seed photon temperature in our fits (see section~\ref{without}), and our best-fit $T_{\rm in}$ values are close to \texttt{Nthcomp} seed photon temperature values measured with \textit{Swift}/XRT \citep{StieleKong2018}. Besides, for all our models, we find the best-fit \texttt{Nthcomp} photon index of $\sim 2$, which is not too different from the values measured with \textit{NuSTAR} \citep{2018ApJ...852L..34X} and \textit{Swift}/XRT \citep{StieleKong2018}.

We note that the \texttt{Nthcomp} best-fit electron temperature ($kT_{\rm e}$) value is $\sim 2.5$~keV, when we fit only the soft X-ray spectrum from the SXT (Table~\ref{tab:bestfit}). A similar small value was inferred from the \textit{Swift}/XRT soft X-ray spectrum \citep{StieleKong2018}. On the other hand, if we fit the broadband spectrum (from SXT+LAXPC), we find higher values ($\sim 20-22$~keV) of $kT_{\rm e}$ for our models M3--M7 (Table~\ref{tab:bestfit}). These values are consistent with the $kT_{\rm e}$ value of $\sim 20$~keV \citep{2018ApJ...852L..34X} measured from the combined soft+hard X-ray spectrum of \textit{NuSTAR}. From these, we conclude that the Comptonization component cannot be correctly characterized by fitting only an observed soft X-ray spectrum for MAXI J1535--571. Moreover, these generally indicate that the analysis of the broadband spectrum can be crucial to characterize the continuum spectral components.
We also find that the \texttt{Nthcomp} photon index and seed photon temperature do not depend much on whether the hard X-ray spectrum is included for fitting or not (Table~\ref{tab:bestfit}). Such an insensitivity of the seed photon temperature, but the sensitivity of $kT_{\rm e}$, to the inclusion of the hard X-ray spectrum is not surprising, as the seed photon temperature is connected to the low energy rollover, while $kT_{\rm e}$ determines the high energy rollover. 

In the spectral model M5, we include the \texttt{Laor} component to describe the relativistic Fe K$\alpha$ emission line. After fitting, this model gives a disc inclination angle $i$ of $\sim 36^\circ - 78^\circ$ (90\% range; Table~\ref{tab:bestfit}). While the $i$ value is not very constrained here, it is consistent with $i$ values measured with \textit{NuSTAR} \citep{2018ApJ...852L..34X} and \textit{NICER} \citep{2018ApJ...860L..28M}. The \texttt{Laor} component also gives an accretion disc inner edge  radius of $\sim 2.3-14.3~R_{\rm g}$ (90\% range; Table~\ref{tab:bestfit}).

In this paper, we also address the reflection features in the spectrum by performing a full-fledged reflection fitting using the physically motivated \textsc{relxill} family of relativistic reflection models, assuming a lamp-post geometry for the Comptonizing cloud (section~\ref{with}). While \cite{2018ApJ...852L..34X} and \cite{2018ApJ...860L..28M} used different flavours of the \textsc{relxill} model to account for the incident continuum and the reflected spectrum by setting the reflection fraction as a free parameter, we rather explicitly include \texttt{Nthcomp} as the source for Comptonized spectra being incident on the disc, and freeze the reflection fraction ($R_{f}$) to -1, thus allowing the reflection models to only return for the relativistically blurred and unblurred reflection components of the spectra.
Considering only the blurred reflection component (model M6; Tables~\ref{tab:models} and \ref{tab:bestfit}), we constrain the height of the corona to $7.3_{-1.1}^{+1.3}~R_\mathrm{g}$, log of the ionization parameter to $3.59_{-0.17}^{+0.32}$ log(erg cm s$^{-1}$) and the inner disc truncation at $<1.61~R_\mathrm{ISCO}$. These values agree with the reflection fitting results using \textit{NuSTAR} \citep{2018ApJ...852L..34X}.
The 90\% range of our best-fit disc inclination angle ($i$) value from the \texttt{relxilllpCp} spectral component of the model M6 is $\sim 35^\circ-47^\circ$, which is also consistent with the best-fit $i$ values from the independent \texttt{Laor} component of our model M5. 
While \textit{NuSTAR} and \textit{NICER} data analyses indicated a high value of the black hole spin parameter $a_{\star}$ \citep{2018ApJ...852L..34X, 2018ApJ...860L..28M}, we find intermediate values of $a_{\star}$ ($0.67_{-0.12}^{+0.11}$).
Note that the ISCO of a black hole with a spin parameter of $a_{\star}=0.67$ is at 3.53 $R_\mathrm{g}$ \citep{1972ApJ...178..347B}. Therefore, our estimate of the upper limit of inner disc truncation radius ($1.61~R_\mathrm{ISCO}$; model: M6) corresponds to 5.68 $R_\mathrm{g}$, which is consistent with the estimate from the independent \texttt{Laor} component (Model M5; Table~\ref{tab:bestfit}), again suggesting the reliability of our results.

As mentioned in section~\ref{with}, by fitting with our model M6, we find an excess in the residual between 6 keV and 9 keV (Figure~\ref{fig:reflection}). This could be due to a narrow Fe K$\alpha$ line complex, which was earlier found with \textit{NuSTAR} and \textit{NICER} \citep{2018ApJ...852L..34X, 2018ApJ...860L..28M}. We, therefore, include a spectral component (in model M7) to describe the reprocessing by distant material. With the model M7, we find that the fit is slightly better (Table~\ref{tab:models}), and there is no major change in best-fit parameter values (relative to model M6), except for the height of the Comptonizing source and the disc inclination angle $i$ (Table~\ref{tab:bestfit}).

In our models M6 and M7, which include one or more full-fledged reflection spectral components, we use the \texttt{Kerrbb} component \citep{2005ApJS..157..335L}, instead of the \texttt{diskbb} component, to describe the multi-colour disc blackbody spectrum. Among all the free parameters of the \texttt{Kerrbb} model\footnote{\url{https://heasarc.nasa.gov/xanadu/XSPEC/manual/XSmodelKerrbb.html}}, the key parameters that we are interested in are, the mass of the black hole, $M_\mathrm{BH}$, distance to the source, and the accretion rate, \.{M}. The \texttt{Kerrbb} model had primarily been used in the context of constraining the spin of the black hole. However, since the model has too many parameters, the standard way has been to freeze some of the parameters (e.g., $M_\mathrm{BH}$, source inclination) to its previously measured values, to constrain the spin. But, since our total models (M6 and M7) have a reflection spectral component with inclination and black hole spin as free parameters, we naturally tie these two parameters of the reflection component with that of the \texttt{Kerrbb} component, and keep $M_\mathrm{BH}$, \.{M} and distance as free parameters, and fit for the same. (Refer \cite{1997ApJ...482L.155Z} and \cite{2005ApJS..157..335L} for more information on the underlying methodology behind the estimation of the aforementioned parameters).

This is the first time, to the best of our knowledge, the \texttt{Kerrbb} component is used for MAXI J1535--571, and we estimate the mass of the central black hole to be $10.39_{-0.62}^{+0.61}~M_{\odot}$, accretion rate to be $\approx 0.31 \times 10^{17}$ g s$^{-1}$ and the source distance to be $5.4_{-1.1}^{+1.8}$ kpc (model M7). Our estimate of the black hole mass is close to that of \cite{Shang2018} ($8.8_{-1.1}^{+1.2}$M$_{\odot}$), who estimated the mass using a two component advective flow (TCAF) solution \citep{1995ApJ...455..623C}. Our distance estimate of $5.4_{-1.1}^{+1.8}$ kpc is also in agreement with that (6.5 kpc) of \cite{2017ATel10711....1R}, who estimated this parameter using the ATCA radio observations at 5.5 and 9 GHz frequencies, assuming the source to be as close to the Galactic centre as possible along its line-of-sight.

With these estimated physical parameters, we also calculate the luminosity at which this source underwent its hard to intermediate state transition. Considering a bolometric flux of $1.6\times 10^{-7}$ erg s$^{-1}$cm$^{-2}$ and an average distance of $\approx 5.3$ kpc (from models M6 and M7), we estimate the transition luminosity to be, $\sim5.4\times 10^{38}$ erg s$^{-1}$. This corresponds to $\sim$45\% $L_\mathrm{Edd}$ for this source, and exceeds the Eddington luminosity for a 1.4$M_{\odot}$ canonical neutron star by a factor of $\sim$3, suggesting, in an independent way, that MAXI J1535--571 indeed hosts a black hole. 

\section*{Acknowledgements}

The authors acknowledge the supports from Indian Space Research Organisation (ISRO) for mission operations and distributions of the data through ISSDC. The LAXPC Payload Operations Center (POC), TIFR, Mumbai is acknowledged for providing us important inputs regarding data analysis and also for the required software tools.
This work has used the data from the Soft X-ray Telescope (SXT) developed at TIFR, Mumbai, and the SXT POC at TIFR is thanked for verifying and releasing the data via the ISSDC data archive and providing the necessary software tools.
This research has made use of the \textit{MAXI} data provided by RIKEN, JAXA and the \textit{MAXI} team. NS acknowledges the support from TIFR-Mumbai and the DST-INSPIRE fellowship. SC acknowledges CSR-NWU, Potchefstroom for supporting his {\it AstroSat} related projects.




\bibliographystyle{mnras}
\bibliography{msv3} 

\begin{thebibliography}{}
\makeatletter
\relax
\def\mn@urlcharsother{\let\do\@makeother \do\$\do\&\do\#\do\^\do\_\do\%\do\~}
\def\mn@doi{\begingroup\mn@urlcharsother \@ifnextchar [ {\mn@doi@}
  {\mn@doi@[]}}
\def\mn@doi@[#1]#2{\def\@tempa{#1}\ifx\@tempa\@empty \href
  {http://dx.doi.org/#2} {doi:#2}\else \href {http://dx.doi.org/#2} {#1}\fi
  \endgroup}
\def\mn@eprint#1#2{\mn@eprint@#1:#2::\@nil}
\def\mn@eprint@arXiv#1{\href {http://arxiv.org/abs/#1} {{\tt arXiv:#1}}}
\def\mn@eprint@dblp#1{\href {http://dblp.uni-trier.de/rec/bibtex/#1.xml}
  {dblp:#1}}
\def\mn@eprint@#1:#2:#3:#4\@nil{\def\@tempa {#1}\def\@tempb {#2}\def\@tempc
  {#3}\ifx \@tempc \@empty \let \@tempc \@tempb \let \@tempb \@tempa \fi \ifx
  \@tempb \@empty \def\@tempb {arXiv}\fi \@ifundefined
  {mn@eprint@\@tempb}{\@tempb:\@tempc}{\expandafter \expandafter \csname
  mn@eprint@\@tempb\endcsname \expandafter{\@tempc}}}

\bibitem[\protect\citeauthoryear{{Agrawal}}{{Agrawal}}{2006}]{2006AdSpR..38.2989A}
{Agrawal} P.~C.,  2006, \mn@doi [Advances in Space Research]
  {10.1016/j.asr.2006.03.038}, \href
  {http://adsabs.harvard.edu/abs/2006AdSpR..38.2989A} {38, 2989}

\bibitem[\protect\citeauthoryear{{Agrawal} et~al.,}{{Agrawal}
  et~al.}{2017}]{2017JApA...38...30A}
{Agrawal} P.~C.,  et~al., 2017, \mn@doi [Journal of Astrophysics and Astronomy]
  {10.1007/s12036-017-9451-z}, \href
  {http://adsabs.harvard.edu/abs/2017JApA...38...30A} {38, 30}

\bibitem[\protect\citeauthoryear{{Antia} et~al.,}{{Antia}
  et~al.}{2017}]{2017ApJS..231...10A}
{Antia} H.~M.,  et~al., 2017, \mn@doi [\apjs] {10.3847/1538-4365/aa7a0e}, \href
  {http://adsabs.harvard.edu/abs/2017ApJS..231...10A} {231, 10}

\bibitem[\protect\citeauthoryear{{Arnaud}}{{Arnaud}}{1996}]{1996ASPC..101...17A}
{Arnaud} K.~A.,  1996, in {Jacoby} G.~H.,  {Barnes} J.,  eds,  Astronomical
  Society of the Pacific Conference Series Vol. 101, Astronomical Data Analysis
  Software and Systems V. p.~17

\bibitem[\protect\citeauthoryear{{Bardeen}, {Press}  \& {Teukolsky}}{{Bardeen}
  et~al.}{1972}]{1972ApJ...178..347B}
{Bardeen} J.~M.,  {Press} W.~H.,   {Teukolsky} S.~A.,  1972, \mn@doi [\apj]
  {10.1086/151796}, \href {http://adsabs.harvard.edu/abs/1972ApJ...178..347B}
  {178, 347}

\bibitem[\protect\citeauthoryear{{Chakrabarti} \& {Titarchuk}}{{Chakrabarti} \&
  {Titarchuk}}{1995}]{1995ApJ...455..623C}
{Chakrabarti} S.,  {Titarchuk} L.~G.,  1995, \mn@doi [\apj] {10.1086/176610},
  \href {http://adsabs.harvard.edu/abs/1995ApJ...455..623C} {455, 623}

\bibitem[\protect\citeauthoryear{{Dauser}, {Garc{\'{\i}}a}, {Parker}, {Fabian}
  \& {Wilms}}{{Dauser} et~al.}{2014}]{2014MNRAS.444L.100D}
{Dauser} T.,  {Garc{\'{\i}}a} J.,  {Parker} M.~L.,  {Fabian} A.~C.,   {Wilms}
  J.,  2014, \mn@doi [\mnras] {10.1093/mnrasl/slu125}, \href
  {http://adsabs.harvard.edu/abs/2014MNRAS.444L.100D} {444, L100}

\bibitem[\protect\citeauthoryear{{Fabian}, {Rees}, {Stella}  \&
  {White}}{{Fabian} et~al.}{1989}]{1989MNRAS.238..729F}
{Fabian} A.~C.,  {Rees} M.~J.,  {Stella} L.,   {White} N.~E.,  1989, \mn@doi
  [\mnras] {10.1093/mnras/238.3.729}, \href
  {http://adsabs.harvard.edu/abs/1989MNRAS.238..729F} {238, 729}

\bibitem[\protect\citeauthoryear{{Garc{\'{\i}}a} \& {Kallman}}{{Garc{\'{\i}}a}
  \& {Kallman}}{2010}]{2010ApJ...718..695G}
{Garc{\'{\i}}a} J.,  {Kallman} T.~R.,  2010, \mn@doi [\apj]
  {10.1088/0004-637X/718/2/695}, \href
  {http://adsabs.harvard.edu/abs/2010ApJ...718..695G} {718, 695}

\bibitem[\protect\citeauthoryear{{Garc{\'{\i}}a} et~al.,}{{Garc{\'{\i}}a}
  et~al.}{2014}]{2014ApJ...782...76G}
{Garc{\'{\i}}a} J.,  et~al., 2014, \mn@doi [\apj] {10.1088/0004-637X/782/2/76},
  \href {http://adsabs.harvard.edu/abs/2014ApJ...782...76G} {782, 76}

\bibitem[\protect\citeauthoryear{{Gendreau} et~al.,}{{Gendreau}
  et~al.}{2017}]{2017ATel10768....1G}
{Gendreau} K.,  et~al., 2017, The Astronomer's Telegram, \href
  {http://adsabs.harvard.edu/abs/2017ATel10768....1G} {10768}

\bibitem[\protect\citeauthoryear{{Harrison} et~al.,}{{Harrison}
  et~al.}{2013}]{2013ApJ...770..103H}
{Harrison} F.~A.,  et~al., 2013, \mn@doi [\apj] {10.1088/0004-637X/770/2/103},
  \href {http://adsabs.harvard.edu/abs/2013ApJ...770..103H} {770, 103}

\bibitem[\protect\citeauthoryear{{Huang} et~al.,}{{Huang}
  et~al.}{2018}]{2018ApJ...866..122H}
{Huang} Y.,  et~al., 2018, \mn@doi [\apj] {10.3847/1538-4357/aade4c}, \href
  {http://adsabs.harvard.edu/abs/2018ApJ...866..122H} {866, 122}

\bibitem[\protect\citeauthoryear{{Kennea}, {Evans}, {Beardmore}, {Krimm},
  {Romano}, {Yamaoka}, {Serino}  \& {Negoro}}{{Kennea}
  et~al.}{2017}]{2017ATel10700....1K}
{Kennea} J.~A.,  {Evans} P.~A.,  {Beardmore} A.~P.,  {Krimm} H.~A.,  {Romano}
  P.,  {Yamaoka} K.,  {Serino} M.,   {Negoro} H.,  2017, The Astronomer's
  Telegram, \href {http://adsabs.harvard.edu/abs/2017ATel10700....1K} {10700}

\bibitem[\protect\citeauthoryear{{Laor}}{{Laor}}{1991}]{1991ApJ...376...90L}
{Laor} A.,  1991, \mn@doi [\apj] {10.1086/170257}, \href
  {http://adsabs.harvard.edu/abs/1991ApJ...376...90L} {376, 90}

\bibitem[\protect\citeauthoryear{{Li}, {Zimmerman}, {Narayan}  \&
  {McClintock}}{{Li} et~al.}{2005}]{2005ApJS..157..335L}
{Li} L.~X.,  {Zimmerman} E.~R.,  {Narayan} R.,   {McClintock} J.~E.,  2005,
  \mn@doi [\apjs] {10.1086/428089}, \href
  {http://adsabs.harvard.edu/abs/2005ApJS..157..335L} {157, 335}

\bibitem[\protect\citeauthoryear{{Mereminskiy}, {Grebenev}, {Prosvetov}  \&
  {Semena}}{{Mereminskiy} et~al.}{2018}]{2018AstL...44..378M}
{Mereminskiy} I.~A.,  {Grebenev} S.~A.,  {Prosvetov} A.~V.,   {Semena} A.~N.,
  2018, \mn@doi [Astronomy Letters] {10.1134/S106377371806004X}, \href
  {http://adsabs.harvard.edu/abs/2018AstL...44..378M} {44, 378}

\bibitem[\protect\citeauthoryear{{Miller} et~al.,}{{Miller}
  et~al.}{2018}]{2018ApJ...860L..28M}
{Miller} J.~M.,  et~al., 2018, \mn@doi [\apjl] {10.3847/2041-8213/aacc61},
  \href {http://adsabs.harvard.edu/abs/2018ApJ...860L..28M} {860, L28}

\bibitem[\protect\citeauthoryear{{Nakahira} et~al.,}{{Nakahira}
  et~al.}{2017}]{2017ATel10729....1N}
{Nakahira} S.,  et~al., 2017, The Astronomer's Telegram, \href
  {http://adsabs.harvard.edu/abs/2017ATel10729....1N} {10729}

\bibitem[\protect\citeauthoryear{{Negoro} et~al.,}{{Negoro}
  et~al.}{2017}]{2017ATel10699....1N}
{Negoro} H.,  et~al., 2017, The Astronomer's Telegram, \href
  {http://adsabs.harvard.edu/abs/2017ATel10699....1N} {10699}

\bibitem[\protect\citeauthoryear{{Negoro} et~al.,}{{Negoro}
  et~al.}{2018a}]{2018ATel11568....1N}
{Negoro} H.,  et~al., 2018a, The Astronomer's Telegram, \href
  {http://adsabs.harvard.edu/abs/2018ATel11568....1N} {11568}

\bibitem[\protect\citeauthoryear{{Negoro} et~al.,}{{Negoro}
  et~al.}{2018b}]{2018ATel11682....1N}
{Negoro} H.,  et~al., 2018b, The Astronomer's Telegram, \href
  {http://adsabs.harvard.edu/abs/2018ATel11682....1N} {11682}

\bibitem[\protect\citeauthoryear{{Novikov} \& {Thorne}}{{Novikov} \&
  {Thorne}}{1973}]{1973blho.conf..343N}
{Novikov} I.~D.,  {Thorne} K.~S.,  1973, in {Dewitt} C.,  {Dewitt} B.~S.,  eds,
  Black Holes (Les Astres Occlus). pp 343--450

\bibitem[\protect\citeauthoryear{{Parker} et~al.,}{{Parker}
  et~al.}{2015}]{2015ApJ...808....9P}
{Parker} M.~L.,  et~al., 2015, \mn@doi [\apj] {10.1088/0004-637X/808/1/9},
  \href {http://adsabs.harvard.edu/abs/2015ApJ...808....9P} {808, 9}

\bibitem[\protect\citeauthoryear{{Protassov}, {van Dyk}, {Connors}, {Kashyap}
  \& {Siemiginowska}}{{Protassov} et~al.}{2002}]{2002ApJ...571..545P}
{Protassov} R.,  {van Dyk} D.~A.,  {Connors} A.,  {Kashyap} V.~L.,
  {Siemiginowska} A.,  2002, \mn@doi [\apj] {10.1086/339856}, \href
  {http://adsabs.harvard.edu/abs/2002ApJ...571..545P} {571, 545}

\bibitem[\protect\citeauthoryear{{Russell}, {Miller-Jones}, {Sivakoff},
  {Tetarenko}  \& {Jacpot Xrb Collaboration}}{{Russell}
  et~al.}{2017}]{2017ATel10711....1R}
{Russell} T.~D.,  {Miller-Jones} J.~C.~A.,  {Sivakoff} G.~R.,  {Tetarenko}
  A.~J.,   {Jacpot Xrb Collaboration} 2017, The Astronomer's Telegram, \href
  {http://adsabs.harvard.edu/abs/2017ATel10711....1R} {10711}

\bibitem[\protect\citeauthoryear{{Scaringi} \& {ASTR211 Students}}{{Scaringi}
  \& {ASTR211 Students}}{2017}]{2017ATel10702....1S}
{Scaringi} S.,  {ASTR211 Students} 2017, The Astronomer's Telegram, \href
  {http://adsabs.harvard.edu/abs/2017ATel10702....1S} {10702}

\bibitem[\protect\citeauthoryear{{Shakura} \& {Sunyaev}}{{Shakura} \&
  {Sunyaev}}{1973}]{1973A&A....24..337S}
{Shakura} N.~I.,  {Sunyaev} R.~A.,  1973, \aap, \href
  {http://adsabs.harvard.edu/abs/1973A\%26A....24..337S} {24, 337}

\bibitem[\protect\citeauthoryear{{Shang}, {Debnath}, {Chatterjee}, {Jana},
  {Chakrabarti}, {Chang}, {Yap}  \& {Chiu}}{{Shang} et~al.}{2018}]{Shang2018}
{Shang} J.~R.,  {Debnath} D.,  {Chatterjee} D.,  {Jana} A.,  {Chakrabarti}
  S.~K.,  {Chang} H.-K.,  {Yap} Y.-X.,   {Chiu} J.-L.,  2018, preprint, \href
  {http://adsabs.harvard.edu/abs/2018arXiv180607147S} {} (\mn@eprint {arXiv}
  {1806.07147})

\bibitem[\protect\citeauthoryear{{Shidatsu} et~al.,}{{Shidatsu}
  et~al.}{2017a}]{2017ATel10761....1S}
{Shidatsu} M.,  et~al., 2017a, The Astronomer's Telegram, \href
  {http://adsabs.harvard.edu/abs/2017ATel10761....1S} {10761}

\bibitem[\protect\citeauthoryear{{Shidatsu} et~al.,}{{Shidatsu}
  et~al.}{2017b}]{2017ATel11020....1S}
{Shidatsu} M.,  et~al., 2017b, The Astronomer's Telegram, \href
  {http://adsabs.harvard.edu/abs/2017ATel11020....1S} {11020}

\bibitem[\protect\citeauthoryear{{Shimura} \& {Takahara}}{{Shimura} \&
  {Takahara}}{1995}]{1995ApJ...445..780S}
{Shimura} T.,  {Takahara} F.,  1995, \mn@doi [\apj] {10.1086/175740}, \href
  {http://adsabs.harvard.edu/abs/1995ApJ...445..780S} {445, 780}

\bibitem[\protect\citeauthoryear{{Singh} et~al.,}{{Singh}
  et~al.}{2016}]{2016SPIE.9905E..1ES}
{Singh} K.~P.,  et~al., 2016, in Space Telescopes and Instrumentation 2016:
  Ultraviolet to Gamma Ray. p. 99051E, \mn@doi{10.1117/12.2235309}

\bibitem[\protect\citeauthoryear{{Singh} et~al.,}{{Singh}
  et~al.}{2017}]{2017JApA...38...29S}
{Singh} K.~P.,  et~al., 2017, \mn@doi [Journal of Astrophysics and Astronomy]
  {10.1007/s12036-017-9448-7}, \href
  {http://adsabs.harvard.edu/abs/2017JApA...38...29S} {38, 29}

\bibitem[\protect\citeauthoryear{{Stiele} \& {Kong}}{{Stiele} \&
  {Kong}}{2018}]{StieleKong2018}
{Stiele} H.,  {Kong} A.~K.~H.,  2018, preprint, \href
  {http://adsabs.harvard.edu/abs/2018arXiv181007203S} {} (\mn@eprint {arXiv}
  {1810.07203})

\bibitem[\protect\citeauthoryear{{Tao} et~al.,}{{Tao}
  et~al.}{2018}]{2018MNRAS.480.4443T}
{Tao} L.,  et~al., 2018, \mn@doi [\mnras] {10.1093/mnras/sty2157}, \href
  {http://adsabs.harvard.edu/abs/2018MNRAS.480.4443T} {480, 4443}

\bibitem[\protect\citeauthoryear{{Tetarenko}, {Russell}, {Miller-Jones},
  {Sivakoff}  \& {Jacpot Xrb Collaboration}}{{Tetarenko}
  et~al.}{2017}]{2017ATel10745....1T}
{Tetarenko} A.~J.,  {Russell} T.~D.,  {Miller-Jones} J.~C.~A.,  {Sivakoff}
  G.~R.,   {Jacpot Xrb Collaboration} 2017, The Astronomer's Telegram, \href
  {http://adsabs.harvard.edu/abs/2017ATel10745....1T} {10745}

\bibitem[\protect\citeauthoryear{{Verner}, {Ferland}, {Korista}  \&
  {Yakovlev}}{{Verner} et~al.}{1996}]{1996ApJ...465..487V}
{Verner} D.~A.,  {Ferland} G.~J.,  {Korista} K.~T.,   {Yakovlev} D.~G.,  1996,
  \mn@doi [\apj] {10.1086/177435}, \href
  {http://adsabs.harvard.edu/abs/1996ApJ...465..487V} {465, 487}

\bibitem[\protect\citeauthoryear{{Wilms}, {Allen}  \& {McCray}}{{Wilms}
  et~al.}{2000}]{2000ApJ...542..914W}
{Wilms} J.,  {Allen} A.,   {McCray} R.,  2000, \mn@doi [\apj] {10.1086/317016},
  \href {http://adsabs.harvard.edu/abs/2000ApJ...542..914W} {542, 914}

\bibitem[\protect\citeauthoryear{{Xu} et~al.,}{{Xu}
  et~al.}{2018}]{2018ApJ...852L..34X}
{Xu} Y.,  et~al., 2018, \mn@doi [\apjl] {10.3847/2041-8213/aaa4b2}, \href
  {http://adsabs.harvard.edu/abs/2018ApJ...852L..34X} {852, L34}

\bibitem[\protect\citeauthoryear{{Zdziarski}, {Johnson}  \&
  {Magdziarz}}{{Zdziarski} et~al.}{1996}]{1996MNRAS.283..193Z}
{Zdziarski} A.~A.,  {Johnson} W.~N.,   {Magdziarz} P.,  1996, \mn@doi [\mnras]
  {10.1093/mnras/283.1.193}, \href
  {http://adsabs.harvard.edu/abs/1996MNRAS.283..193Z} {283, 193}

\bibitem[\protect\citeauthoryear{{Zhang}, {Cui}  \& {Chen}}{{Zhang}
  et~al.}{1997}]{1997ApJ...482L.155Z}
{Zhang} S.~N.,  {Cui} W.,   {Chen} W.,  1997, \mn@doi [\apjl] {10.1086/310705},
  \href {http://adsabs.harvard.edu/abs/1997ApJ...482L.155Z} {482, L155}

\bibitem[\protect\citeauthoryear{{{\.Z}ycki}, {Done}  \& {Smith}}{{{\.Z}ycki}
  et~al.}{1999}]{1999MNRAS.309..561Z}
{{\.Z}ycki} P.~T.,  {Done} C.,   {Smith} D.~A.,  1999, \mn@doi [\mnras]
  {10.1046/j.1365-8711.1999.02885.x}, \href
  {http://adsabs.harvard.edu/abs/1999MNRAS.309..561Z} {309, 561}

\makeatother
\end{thebibliography}



\appendix
\section{Pile-up check}\label{appendix:pile-up}
The aim of this appendix is to check if a plausible pile-up in the SXT could significantly affect our final results and conclusions. Since the conclusions are based on the fitting of SXT+LAXPC data, in order to check for the pile-up effect, we use the same combined SXT+LAXPC data, as used in our analysis. We fit the model M4 to four different groups of SXT+LAXPC data, each group differing from one another by the area of the central region excluded from the source PSF on the SXT CCD.

\begin{table}
\caption{Best fit parameters of the model M4 (see Table~\ref{tab:models}) from the fit to the SXT+LAXPC data of MAXI J1535--571, for different sizes of the excluded inner region in the source PSF on the SXT CCD.}
 \label{tab:pileup}
\scalebox{0.9}{
 \hspace{-0.7cm}
 \begin{tabular}{cccccc}
   
   \hline
   \hline
  Removal & $N_{\rm H}$ & $T_{\rm in}$ & $\Gamma$ & $kT_{\rm e}$ & $\chi^2_{\nu}$\\  
  radius & $(\times 10^{22}$cm$^{-2})$ & (keV) &  & (keV) & \\  
   \hline
  
0' & $2.92_{-0.25}^{+0.17}$ & $0.26_{-0.03}^{+0.02}$ & $2.28_{-0.02}^{+0.01}$ & $21.2_{-1.3}^{+1.5}$ & 1.063 \\
 0.5' & $2.91_{-0.14}^{+0.23}$ & $0.28\pm0.01$ & $2.29\pm0.01$ & $23.1_{-1.5}^{+1.6}$ & 1.056 \\
 1.0' & $2.87_{-0.15}^{+0.26}$ & $0.25\pm0.01$ & $2.30\pm0.01$ & $23.2_{-1.6}^{+1.7}$ & 1.011 \\
 1.5' & $2.90_{-0.12}^{+0.12}$ & $0.29\pm0.01$ & $2.30\pm0.01$ & $23.8_{-1.5}^{+2.1}$ & 0.929 \\
 2.0' & $2.87_{-0.11}^{+0.23}$ & $0.28\pm0.01$ & $2.31\pm0.01$ & $24.3_{-1.9}^{+1.7}$ & 0.882 \\

   \hline
 \end{tabular}}
\begin{flushleft}
 \textbf{Note:} The first column represents the radius of the central region (in arcmin) excluded from the source PSF on the SXT CCD; the second column is the hydrogen absorption column density ($n_{\rm H}$); third column is the inner disc temperature ($T_{\rm in}$); fourth column is the asymptotic power-law photon index ($\Gamma$); fifth column is the coronal electron temperature ($kT_{\rm e}$); and the sixth column is the reduced $\chi^2$ to the fit, defined as the ratio of $\chi^2$ to the number of degrees of freedom.
\end{flushleft}
\end{table}

This model is chosen, as it is simple enough to fit the data with lesser photon counts (in the cases of larger excluded central region), yet describes the disc and coronal parameters using physical model components and the Fe line by a Gaussian proxy. Best fit parameters obtained from fitting M4 to four different groups of SXT+LAXPC data are displayed in Table \ref{tab:pileup}. Only those parameters that sufficiently describe a systematic evolution of the spectral hardness, if seen, with respect to change in annular radii are displayed. It can be seen from Table \ref{tab:pileup} that, a plausible SXT pile-up is not significant enough to change the conclusions of this paper. Therefore, we do not exclude any portion of the SXT PSF, in order to have a good photon count statistics, which is also required for fitting the data with the complex reflection models.


\bsp	
\label{lastpage}
\end{document}